# Transport and infrared properties of SmFeAs($O_{1-x}F_x$): from SDW to superconducting ordering

M. Tropeano, C. Fanciulli, C. Ferdeghini, D. Marrè, A.S. Siri, M. Putti
*CNR-INFM-LAMIA and Dipartimento di Fisica, Via Dodecaneso 33, 16146 Genova, Italy*

A. Martinelli, M. Ferretti, A. Palenzona
*CNR-INFM-LAMIA and Dipartimento di Chimica e Chimica Industriale, Via Dodecaneso 31, 16146 Genova, Italy*

M. R. Cimberle
*CNR-IMEM, Dipartimento di Fisica, Via Dodecaneso 33, 16146 Genova, Italy*

C. Mirri, S. Lupi, R. Sopracase, P. Calvani
*CNR-INFM "Coherentia" and Dipartimento di Fisica, Università di Roma "La Sapienza", P.le A. Moro 2, 00185 Rome (Italy)*

A. Perucchi
*Sincrotrone Trieste S.C.p.A.,S.S. 14 km 163.5, in Area Science Park, 34012 Basovizza, Trieste, Italy*

**Abstract**
We report measurements of resistivity, magnetoresistivity, Hall effect, Seebeck coefficient, infrared reflectivity of undoped SmFeAsO and lightly doped SmFeAs($O_{0.93}F_{0.07}$) oxypnictides. All the properties measured on SmFeAsO are characterized by clear signatures of the magnetic instability. A self-consistent picture emerges in which below the magnetic transition carrier condensation occurs due to the opening of spin density wave (SDW) gap. This is accompanied by the mobility increase of not gapped carriers due to the suppression of electron-electron scattering. SmFeAs($O_{0.93}F_{0.07}$) exhibits an increase of the metallic character on cooling consistent with electron doping, even though at room temperature values of all the properties nearly overlaps with those of SmFeAsO. However, with temperature decrease all anomalies related to the SDW instability are missed and the superconducting transition occurs. This suggests that doping breaks abruptly the symmetries of the Fermi surface inhibiting the SDW formation in favor of the superconducting transition, with no substantial changes in the density of states or in the effective mass.

**Introduction**
The recent discovery of superconductivity at 26 K in LaFeAs($O_{1-x}F_x$) [1] attracted a lot of attention on rare earth (*RE*) oxypnictides. These phases crystallize in the tetragonal system at room temperature and their structure is built up by two kinds of planar layers constituted of edge sharing tetrahedra stacked along the *c*-axis. The former layer is constituted of tetrahedra centred by O with the *RE* at vertices (charge reservoir layer), whereas in the latter Fe coordinates As (conducting layer). The parent compounds exhibit an antiferromagnetic (AF) transition around 140-150 K [1,2,3,4,5,6,7] attributed to the development of a spin density wave (SDW) with a small moment in the Fe-As plane, as indicated by neutron diffraction analysis [3] and by magnetisation measurements [8]. The magnetic transition has been characterized by different experimental techniques [2,9,10,11] and it has been related to a tetragonal - orthorhombic structural transition detected by different techniques at about the same temperature [3,12,13].
Electron doping suppresses the magnetic instability in favour of superconductivity and critical temperature as high as 55 K in the SmFeAs($O_{1-x}F_x$) has been obtained [14].

The layered structure and the rather high critical temperature, hardly explainable by the electron-phonon coupling, seems to suggest a similarity with high temperature superconductors. However, differently from cuprates, many experiments suggest a multi-band nature of superconductivity in these compounds [15,16] as in the case of magnesium diboride.

For the better comprehension of the superconducting mechanism in oxypnictides, a systematic study of electrical and thermoelectrical transport properties and of their dependence upon doping in single crystals would be highly desirable. Unfortunately, up to now, only small single crystals are available [17] and such kind of analysis has been performed mainly on polycrystalline samples [18,19,20], especially by comparison of magnetoresistivity and Hall effect.

In this work we report resistivity, Hall effect, magnetoresistivity, Seebeck coefficient and infrared (IR) conductivity measurements on SmFeAsO and SmFeAs($O_{0.93}F_{0.07}$). The undoped sample exhibits clear anomalies at ~130 - 140 K. Although the nature of this anomaly is not yet definitively explained, in the following we refer to it as due to the occurrence a SDW ordered state. The same properties measured on SmFeAs($O_{0.93}F_{0.07}$) show that low level of doping does not substantially modify phonon and electron parameters, yet it completely suppresses the SDW transition in turn of superconductivity. Concerning the superconducting state, IR reflectivity data show two different spectral features. The former is related to the gap in the *ab*-plane the latter to an interplane Joshepson coupling. These features are similar to those observed in high-Tc cuprates.

**Experimental**

Samples with nominal compositions SmFeAsO and SmFeAs($O_{0.93}F_{0.07}$) were prepared in three steps as reported in [21]: first, heating Sm and As in an evacuated glass flask at a maximum temperature of 550°C, to synthesize SmAs, and then reacting the arsenide with stoichiometric amounts of Fe, $Fe_2O_3$ and $FeF_2$ in a form of a pellet at 1200°C for 24h in an evacuated quartz flask. Finally, the products underwent a further sintering step at 1300°C for 72h in an evacuated quartz flask in order to obtain a compact sample suitable for transport measurements. The effect of sintering is to increase density and connection between the grains, improving substantially transport properties.

Phase identification was performed by XRPD (PHILIPS PW1830; Bragg-Brentano geometry; CuK$_{\alpha1,\alpha2}$; range 20 – 110° $2q$ ; step 0.025° $2q$; sampling time 12 s) and structural refinement was successfully carried out in the space group *P4/nmm* according to the Rietveld method using the program FullProf. The samples were characterised also by scanning electron microscope (SEM) and transmission electron microscope (TEM) analyses. By means of synchrotron powder diffraction, coupled with Rietveld refinement, an orthorhombic to monoclinic phase transition has been ascertained in SmFeAsO, by measurements done at room temperature and 100 K, respectively [13]. Rietveld refinement of X-ray diffraction patterns and SEM observation reveal the single-phase nature of the samples, whereas SEM-EDS analyses are in good agreement with the expected nominal SmFeAsO composition. No evidence for nanodomains, twinning, extended defects or superlattice reflections, can be obtained by TEM observation, thus suggesting a high degree of crystallinity of our samples [21]. The effect of sintering is to increase density and connection between the grains improving substantially transport properties. The good quality of the samples is proved also by thermal [22] and magnetization [8] measurements which are reported elsewhere.

Magnetoresistivity, Hall effect and Seebeck effect were measured by Quantum Design PPMS with Thermal Transport Option from 5 to 300 K in magnetic field up to 9.

Normal incidence reflectivity measurements were performed at the SISSI infrared beamline of the ELETTRA Storage Ring (Trieste), between 10 and 30000 cm$^{-1}$, at temperatures *T* ranging from 5 to 300 K on both samples.

An *in situ* evaporation technique was used to measure the reference. The real part of the optical conductivity $s_1(w)$ was then determined through Kramers-Kronig (KK) transformations and by standard extrapolations of $R(w)$ both at high and low frequency. Details on the experimental technique and data analysis were reported elsewhere [23].

**Resistivity, Hall effect, Hall mobility**

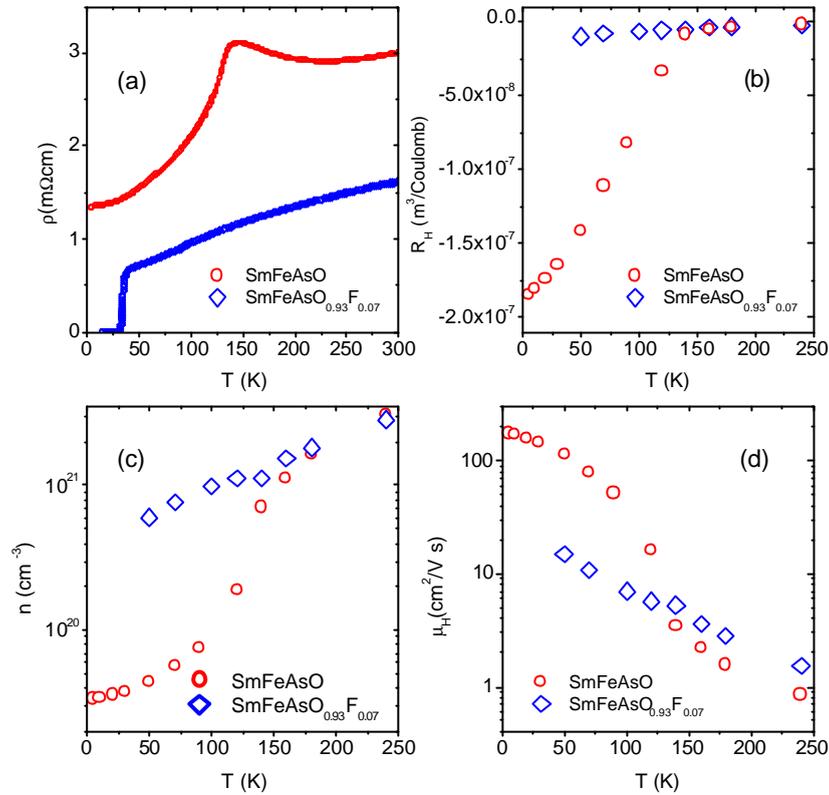

Figure 1. Resistivity and Hall effect measurements for the undoped and 7% F-doped samples: (a) resistivity versus temperature, (b) temperature dependence of the Hall coefficient, (c) carrier density versus temperature and (d) temperature dependence of the Hall mobility.

Resistivity measurements for both samples are compared in figure 1(a). The undoped specimen exhibits a pronounced anomaly around $T\sim135$ K-140 K, then the resistivity decreases monotonically with decreasing temperature. In the following we indicate the temperature at which the resistivity anomaly occurs as $T_{SDW}$. SmFeAs($O_{0.93}F_{0.07}$) is characterized by a nearly linear decrease of resistivity with temperature, with the onset of superconductivity occurring at 36 K.

Hall effect measurements are performed at fixed temperature by sweeping the magnetic field from -9 T to 9 T. A nearly linear dependence of the transverse resistivity on the magnetic field is observed at all the temperatures, differently from what reported for LaFeAs($O_{1-x}F_x$) [18]. The Hall resistance, $R_H$, is shown in fig. 1 (b). It is negative for both the samples, indicating that the main contribution to the Hall effect is electron-like. The electron density, $n$, of the two samples is plotted in fig. 1 (c). The carrier density of the undoped sample strongly depends on temperature showing a stiff drop in correspondence of the resistivity anomaly. This was previously reported in LaFeAsO [11] and was discussed in terms of charge carriers localization at the structural transition. The electron condensation due to the opening of the SDW gap would produce the same features.

For the 7% F doped sample, $n$ varies with temperature in nearly logarithmic way, but no anomaly is observed below $T_{SDW}$. Interestingly the electron densities of the two samples with increasing temperature above $T_{SDW}$ tend to overlap and the values nearly coincide at 250 K. Similar results have been obtained on the same compound in ref. [20] and, on LaFeAs($O_{1-x}F_x$), in ref. [18].

The Hall mobility of the two samples evaluated as $\mu_H = R_H / \rho$ is plotted in figure 1(d). In SmFeAsO $\mu_H$ strongly increases by two orders of magnitude below $T_{SDW}$ and reaches a value of about 200 cm$^2$/V s at 5 K. In ref. [18] a similar behaviour with rather lower values was reported for LaFeAsO. Below T$_{SDW}$, within a SDW framework, charge carriers which are gapped out disappear meanwhile the mobility of the carriers which do not condensate in the SDW state abruptly raises. This suggests that electron-electron scattering is the mechanism which mainly limits the carrier mobility above the SDW transition.

The drop of the resistivity below T$_{SDW}$ can be taken into account by an increased mobility of the carriers which do not participate in the SDW state despite the strongly reduction of carriers density.

The mobility of SmFeAs(O$_{0.93}$F$_{0.07}$) decreases with increasing temperature and, also in this case, there is no evidence of the SDW transition. Interestingly, above T$_{SDW}$, µ$_H$ in the doped sample is larger than in the undoped one. This follows straightforwardly by the lower resistivity values and the nearly equal carrier density of the former sample in comparison with the latter one. Even if, in multiband conduction, µ$_H$ is only an effective quantity, these results suggest that doping does not strongly modify the electronic structure; similar results come out from the analysis of the thermal properties [22] which suggests that the density of states is only slightly modified by F doping. On the contrary, theoretical calculations [24] predict that doping moves the Fermi level into a region of heavier carriers. Our results, suggesting nearly unchanged effective mobility upon doping, disagree with this scenario.

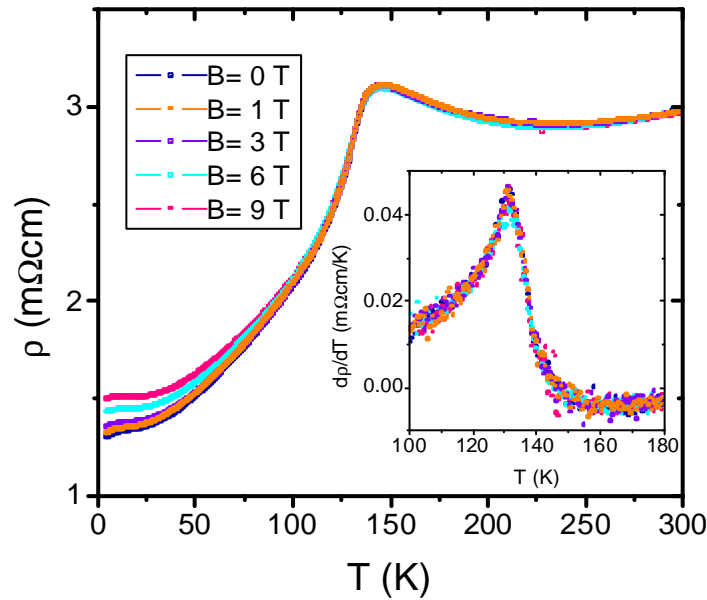

Figure 2. Magnetoresistivity of the undoped sample. In the inset dρ/dT, which underlines T$_{SDW}$

**Magnetoresistivity**

Magnetoresistivity measurements can give further hints on the actual mobility of the two samples. Magnetoresistivity of doped sample reported in ref.[21] is quite low indicating low mobility carriers. It can be roughly estimated of the order of 1% at 50 K and 9 T. Much richer phenomenology is presented by parent compound, whose magnetoresistivity curve as a function of temperature and fixed magnetic field are plotted in fig. 2. A remarkable positive magnetoresistance is visible at low temperature which progressively disappears with increasing temperature towards T$_{SDW}$ as reported also for LaFeAsO [2,11] this is even more evident in figure 3(a) where $\Delta r / r(0) = [r(B) - r(0)] / r(0)$, measured at fixed temperature with increasing magnetic field, is plotted. At 5K and 9T Δρ/ρ(0) is more than 15% and with increasing temperature decreases down to 1% at 120 K. Δρ/ρ(0) measured at higher selected temperatures (140, 160, 180, 240 and 300 K) suddenly drops down to 0.1%.

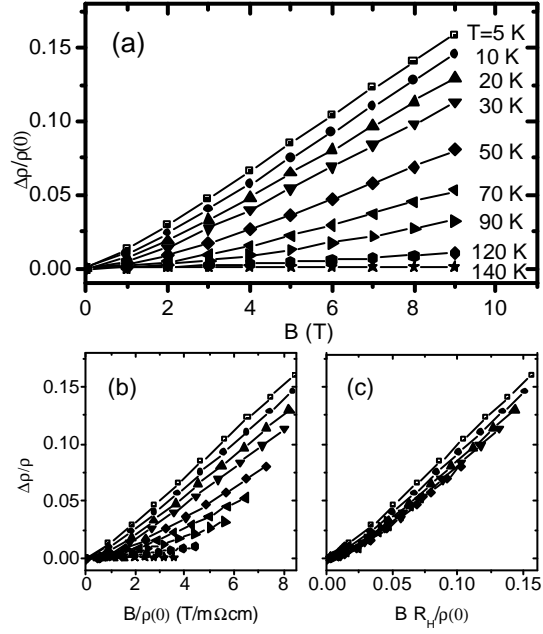

Figure 3. (a) $\Delta r/r(0) = [r(B) - r(0)]/r(0)$, measured at fixed temperature with increasing B. (b) Kholer's plot. (c) $\Delta\rho/\rho(0)$ plotted as a function of $BR_H/r(0) = Bm_H = B\tau/m^*e = \omega_c\tau$ (see text).

We can further notice that the application of the magnetic field does not affect at all the SDW transition. As shown in the inset of fig. 2 dρ/dT, which underlines $T_{SDW}$, does not present any shift and/or enlargement due to the application of the magnetic field.

The temperature dependence of magnetoresistance can be accounted for by the Kholer's rule which assumes Δρ/ρ(0) as a function of the product Bτ where τ is the relaxation time: $\Delta r/r(0) = f(B\tau)$. The relaxation time τ is related to the resistivity by $r = m^*/ne^2\tau$, where m* is the effective mass. If the factor $m^*/ne^2$ does not change with temperature Kholer's rule can be written in the more common form $\Delta r/r(0) = f(B/r(0))$. For SmFeAsO the electron density $n$ is far to be constant below $T_{SDW}$, rather it varies within two order of magnitude as shown in fig.1b. This avoids the usual scaling Δρ/ρ(0) vs B/ρ to work, as evident in fig. 3b. In fig. 3c, Δρ/ρ(0) is plotted as a function of $BR_H/r(0) = Bm_H = B\tau/m^*e = \omega_c\tau$, where $\omega_c$ is the cyclotronic frequency. In this case all the curves collapse together showing that the general Kholer's rule captures the main features of magnetoresistance in a quite extended range of $\omega_c\tau$ (0–0.16). Small differences between the curves, within of ten percent overall, can be appreciated looking at the fig. 3c. Such deviations can be attributed to the presence of more bands that participate in the conduction. Moreover, with varying temperature, scattering mechanisms which contribute differently to magnetoresistance, can come into play. For instance, it was emphasized that below 20 K the resistivity is affected by the ordering of $Sm^{3+}$ sublattice[8,21,22]; in such a case magnetic field might suppress spin fluctuation giving a negative contribution to magnetoresistance which sum up with the positive cyclotronic contribution.

**Seebeck effect**

The Seebeck coefficient, S, of SmFeAsO and SmFeAs($O_{0.93}F_{0.07}$) is shown in fig. 4.

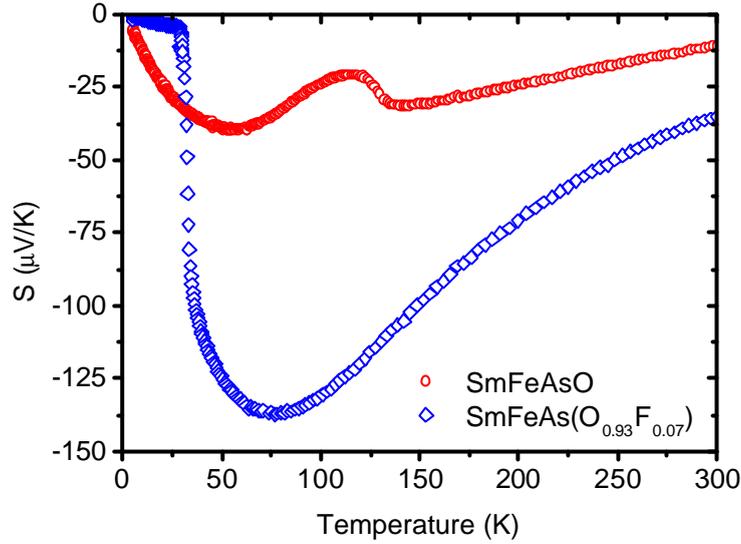

Figure 4. Seebeck coefficient of SmFeAsO and SmFeAs($O_{0.93}F_{0.07}$) versus temperature.

In both samples $S$ is negative over the entire temperature range, indicating that electrons dominate the electrical conduction. This is consistent with the measured negative Hall coefficient (Figure 1(b)).

The two curves present clear signatures of the different electronic transition occurring in the two samples. In the undoped sample, $S$ presents a decrease of its absolute value below $T_{SDW}$; similar behaviour has been observed in LaFeAsO [11].

Anomalies like this have been observed below magnetic instabilities [25,26,27] and can be understood within a free electron model where $S$ is given by:

$$S(T) = -\frac{\pi^2}{3}\frac{k_B}{|e|}k_B T\left[\frac{N(E_F)}{n} + \frac{1}{\tau}\frac{d\tau}{dE_F}\right] \quad (1)$$

where N(E) is the density of states, $E_F$ is the Fermi energy and $\tau$ is the relaxation time. The first term scales as $1/E_F$ and it does not change abruptly at the transition, while the second one, which can be qualitatively related with the changes of mobility with doping and temperature (see fig. 1d), gives the main contribution below the transition. Below $T_{SDW}$, indeed, $\mu_H$ strongly decreases with electron doping suggesting that $d\tau/dE_F$ is large and negative. Above the transition $\mu_H$ is roughly constant with doping and the second term becomes negligible.

In SmFeAs($O_{0.93}F_{0.07}$) $S$ drops to zero below 34 K in reasonable correspondence with the resistive superconducting transition. Actually, it does not reach zero, but it shows a tail which extends till to the lowest temperature, indicating that a minor amount of the sample does not attain the superconducting state.

Neglecting the anomalies related to the ordering transitions, $|S|$ of the two samples has a similar overall behaviour, showing a large maximum around 60-70 K. The values are more than three times higher in the doped than in undoped samples, in agreement with what observed in LaFeAsO [11] and in LaFeAs($O_{0.89}F_{0.11}$) [28]. Looking at eq. (1) it should be the opposite, since $S$ decreases its absolute values with increasing $E_F$. This can be explained considering the multi band nature of the oxypnictides. For two bands, one electron- and the other hole-like, $S$ becomes:

$$S = \frac{\sigma_h|S_h| - \sigma_e|S_e|}{\sigma_h + \sigma_e} \quad (2)$$

Where $\sigma_{e(h)}$ and $S_{e(h)}$ are the contributions of electrons (holes) to the electrical conductivity and Seebeck coefficient, respectively. Looking at eq. (2), the undoped sample might show smaller $S$ values since large but nearly equal hole and electron contributions compensate each-others, whereas electron doping makes the electron contribution emerge.

**Infrared riflectivity**

The far-infrared $R(\omega)$ is shown in Figs.5a and 5c at selected $T$'s for SmFeAsO and SmFeAs($O_{0.93}F_{0.07}$), respectively. R($\omega$) is also reported in the whole spectral range at 300 K in the insets of the same Figures. The reflectivity of both samples resembles that of a bad metal, showing a weak $T$-dependence in the infrared and strong phonon peaks (that will be discussed in a forthcoming paper [29]) at approximately 102, 260, 270, 375 and 450 cm$^{-1}$. Calculations [29] show that the peaks observed at 102, 260 and 450 cm$^{-1}$ correspond to optical phonons polarized along the $c$-axis, while those at 270 and 375 cm$^{-1}$ (which splits in two components at low $T$) are in the $ab$ plane. The presence of both polarizations is in agreement with the polycrystalline nature of the samples. The comparison of our data with some optical $ab$-plane results recently appeared on (Ba$_{1-x}$K$_x$)Fe$_2$As$_2$ [30] and LaFePO [31] single crystals and with ellipsometric data obtained on polycrystalline materials [32] show that the main contribution to the optical properties of pellets comes from the more insulating $c$-axis, whose optical properties have never been measured to our knowledge in the pnictides.

As discussed previously, the electronic and magnetic properties of the undoped material are affected by a SDW instability of the underlying Fermi surface. This transition determines a strong change in the $T$-dependence of the resistivity (see fig.1a): $\rho$ is nearly constant for T>135 K, rapidly decreasing at lower $T$.

The SDW transition affects also the infrared spectrum of the undoped sample, which shows for T<T$_{SDW}$ (in agreement with similar measurements on LaFeAsO [2]) a suppression of R($\omega$) between 250 and 150 cm$^{-1}$ and a more pronounced metallic behavior below the lowest-energy phonon absorption (see fig. 5a). This suppression could not be observed (fig. 5c) in the doped material. Here, due to the absence of the SDW instability, R($\omega$) is more metal-like and shows a monotonic increase at any $T$ for $\omega \rightarrow 0$.

The effect of the SDW transition can be tracked in $\sigma_1(\omega)$, which is plotted for the undoped and the doped sample in Figs. 5b and 5d, respectively. At variance with conventional SDW materials like (TMTSF)$_2$PF$_6$ [33], the transition does not open a gap in the electronic excitations in agreement with transport results (see above) but it induces just a small depression in $\sigma_1(\omega)$ between 250 and 150 cm$^{-1}$. This depression can be better observed in the inset of fig.5b where the difference $\Delta \sigma_1(\omega) = \sigma_1(\omega,T) - \sigma_1(\omega,150K)$ is shown at selected $T$<150 K, and corresponds to a transfer of spectral weight (SW) from high frequencies to those below 100 cm$^{-1}$. The agreement between the dc results (see above) and the IR data across T$_{SDW}$ suggests that the SDW transition partially gaps the Fermi surface. Therefore, the metallic term observed at very low frequency is due to the ungapped states around E$_F$. Its increase below T$_{SDW}$ can be associated with a reduction of the electron-electron scattering.

In the superconducting SmFeAs(O$_{0.93}$F$_{0.07}$) $\sigma_1(\omega)$ is still strongly influenced by the $c$-axis conductivity showing phonon absorptions similar to those observed in the undoped material. However $\sigma_1(\omega)$ does not show any imprint of the SDW transition, in agreement with the linear decrease of the resistivity and with the other transport data (see above). It monotonically increases at any $T$ below 600 cm$^{-1}$.

The effect of the superconducting transition can be directly observed in the inset of fig.5d, where the ratio R($\omega$,T)/R($\omega$,40 K) is reported at different $T$'s. Therein, R($\omega$,T) is the reflectivity in the superconducting state at selected $T$<T$_c$ and R($\omega$,40 K) is that of the normal state. The well evident maximum around 30 cm$^{-1}$, in agreement with similar data collected on polycrystalline

(Nd,Sm)FeAs($O_{0.82}F_{0.18}$) [32] can be associated with the c-axis superconducting (SC) response. Indeed, if in the normal-state the *c*-axis is nearly semiconducting, below $T_c$ it becomes metallic due to the Joshepson interplane coupling. Therefore the low-frequency $R(\omega,T)/R(\omega,40 K)$ allows one to evaluate the *c*-axis Joshepson plasma frequency, which turns out to be about 30 $cm^{-1}$.

At about 100 $cm^{-1}$, $R(\omega,T)/R(\omega,40 K)$ becomes close to 1 as the *c*-axis phonon in this spectral region in practically independent of T. Therefore the two deep minima centered around 75 and 125 $cm^{-1}$ can be interpreted in terms of a single broad minimum with an onset at about 200 $cm^{-1}$. Previous studies on cuprates have shown that this onset approximately indicates the maximum value of the *ab*-plane SC optical gap. With $2\Delta \sim 200$ $cm^{-1}$ one has $2\Delta/k_BT_c \sim 8$, much larger than the BCS value 3.53 which holds for weak coupling. Comparable gap amplitudes are reported for the high-$T_c$ cuprates, where they point toward a strong-coupling SC pairing mechanism. However recent photoemission data on $(Ba_{1-x}K_x)Fe_2As_2$ [34] have shown that two different gaps open below $T_c$. The largest one (*D* ~ ?100 $cm^{-1}$) agrees with the present IR data, while the smallest one (*D* ~ ? 50$cm^{-1}$) has not been observed up to now by optical measurements. In the latter case one would have $2D/k_BT_c \sim 4$ in much better agreement with other experimental techniques.

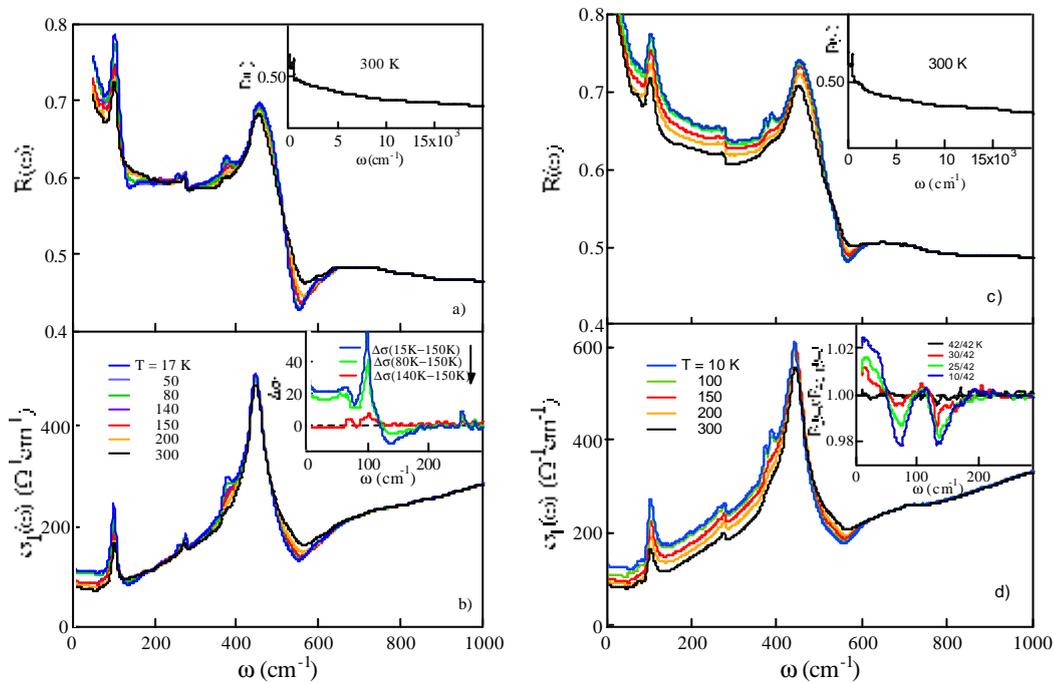

Figure 5. Far-infrared reflectivity (a) and optical conductivity (b) at selected *temperatures* for the undoped SmOFeAs sample. The inset of a) shows the reflectivity at 300 K between 50 and 20000 $cm^{-1}$, that of b) the difference $\Delta\sigma_1=\sigma_1(\omega,T)-\sigma_1(\omega,150 K)$. *c)* and d): same as a) and b), respectively, for the superconducting SmFeAs($O_{0.93}F_{0.07}$) sample. In the inset the ratio *R(w,T)/R(w,42 K)* (see text) is also reported.

**Conclusions**

We report measurements of several transport properties (resistivity, magnetoresistivity, Hall effect, Seebeck coefficient, IR reflectivity ) of undoped SmFeAsO and lightly doped SmFeAs($O_{0.93}F_{0.07}$) oxypnictides. The main purpose was to investigate the effect of the SDW transition on their transport properties and to look for some memory of it at low level of doping.

All the properties measured on SmFeAsO show clear signatures of the magnetic instability. Resistivity measurements show a maximum at about 135 K followed by a sharp drop below 130 K. Hall effect measurements demonstrate that below this temperature the number of carriers abruptly decreases as a consequence of carrier condensation due to the opening of a SDW gap at the Fermi level. Within this framework, the reduction in the resistivity can be explained with an increased mobility of the carriers which remain free, suggesting a strong electron-electron scattering. The presence of high-mobility carriers in the SDW ordered state is fully supported by the measured magnetoresistivity which is rather large at low temperature and drastically reduces with increasing temperature up to $T_{SDW}$. This behavior is consistent with the Kohler rule, only if the temperature dependence of the carrier density is considered.

Also the Seebeck coefficient of SmFeAsO exhibits a clear signature of the SDW transition: below $T_{SDW}$, $S$ changes slope, indicating that a contribution with opposite sign adds to the main term. This additional term arises by the abrupt change of the carrier scattering rate below $T_{SDW}$. Finally, the effect of the SDW transition can be tracked in the IR response. At variance with conventional SDW materials, the ordering transition does not open a gap in the electronic excitations. It induces just a small depression in $\sigma_1(\omega)$ between 250 and 150 cm$^{-1}$, suggesting that the transition partially gaps the Fermi surface. Therefore, the metallic term observed at very low frequency is due to the ungapped states around $E_F$ and its increase below $T_{SDW}$ can be associated with a reduction of the electron-electron scattering.

Concerning the doped sample, transport properties of SmFeAs($O_{0.93}F_{0.07}$) present a more metallic behavior consistent with electron doping, even if the room-temperature values of all the considered quantities nearly overlap in the two samples. This result indicates that F doping does not produce substantial changes in the density of states, nor in the effective mass, as suggested also by thermal properties [10]. However, with decreasing temperature the rich and self-consistent phenomenology summarized up to now completely disappears: all the afore mentioned anomalies are missed, in favor of the occurrence of superconducting transition at around 34 K. Here, the effect of this latter has been also directly observed in the *far-infrared* reflectivity. This allowed us to evaluate the superconducting gap of SmFeAs($O_{0.93}F_{0.07}$), which points toward a pairing mechanism governed by a strong-coupling regime.

## Acknowledgments

The authors thank C. Rizzuto for useful discussion. This work is partially supported by MIUR under the projects PRIN2006021741 and by Compagnia di San Paolo.

## References

[1] Kamihara Y, Watanabe T, Hirano M and Hosono H 2008 *J. Am. Chem. Soc.* **130** 3296
[2] Dong J, Zhang H J, Xu G, Li Z, Li G, Hu W Z, Wu D, Chen G F, Dai X, Luo J L, Fang Z and Wang N L *Europhysics Letters*, 83, 27006 (2008)
[3] de la Cruz C *et al.*, 2008 *Nature* **453** 899
[4] Nakai Y, Ishida K, Kamihara Y, Hirano M and Hosono H 2008 *J. Phys. Soc. Jap.* **77**
[5] Kitao S *et al.* 2008 *Preprint* 0805-0041 [cond-mat]
[6] Klauss H.-H *et al.*, 2008 *Preprint* 0805.0264v1 [cond-mat]
[7] Carlo J.P. *et al.*, 2008 *Preprint* 0805.2186v1 [cond-mat]
[8] Cimberle M R, Ferdeghini C, Canepa F, Ferretti M, Martinelli A, Palenzona A, Siri A S and Tropeano M 2008 *Preprint* 0807.1688 [cond.mat]
[9] H.-H. Klauss, H. Luetkens, R. Klingeler, C. Hess, F. J. Litterst, M. Kraken, M. M. Korshunov, I. Eremin, S.-L. Drechsler, R. Khasanov, A. Amato, J. Hamann-Borrero, N. Leps, A. Kondrat, G. Behr, J. Werner, and B. Buchner, *Phys. Rev. Lett.* 101, 077005 (2008)
[10] Tropeano M, Martinelli A, Palenzona A, Bellingeri E, Galleani dAgliano E, Nguyen T D, Affronte M and Putti M, 2008 *Preprint* 0807.0719 [cond-mat] Phys. Rev. B in press
[11] Mc Guire M A, Christianson A D, Sefat A S, Sales B C, Lumsden M D, Jin R, Payzant E A, Mandrus D, Luan Y, Keppens V, Varadarajan V, Brill J W, Hermann R P, Sougrati M T, Grandjean F and Long G J 2008 *Preprint* 0806.3878v2 [cond-mat]

[12] Nomura T. *et al.*, 2008 *Preprint* 0804.3569 [cond-mat]
[13] Martinelli A, Palenzona A, Ferdeghini C, Putti M and Emerich E 2008 *Preprint* 0808.1024 [cond-mat]
[14] Ren Z A *et al.* 2008 *Chin. Phys. Lett* **25** 2215
[15] Hunte F, Jaroszynski J, Gurevich A, Larbalestier D C, Jin R, Sefat A S, McGuire M A, Sales B C, Christen D K and Mandrus D 2008 *Nature* **453** 903
[16] Zhu X, Yang H, Fang L, Mu G and Wen H-H 2008 *Supercond. Sci. Technol.* **21** 105001
[17] Cheng P et al 2008 *Preprint* 0806.1668v4 [cond-mat]
[18] Kohama Y, Kamihara Y, Baily S A, Civale L, Riggs S C, Balakirev F F, Atake T, Jaime M, Hirano M and Hosono H 2008 *Preprint* 0809.1133 [cond-mat]
[19] Liu R H 2008 *Preprint* 0804.2105v2 [cond-mat]
[20] Liu R H, Wu G, Wu T, Fang D F, Chen H, Li S Y, Liu K, Xie Y L, Wang X F, Yang R L, He C, Feng D L and Chen X H 2008 *Phys. Rev. Lett.* **101** 087001
[21] Martinelli A, Ferretti M, Manfrinetti P, Palenzona A, Tropeano M, Cimberle M R, Ferdeghini C, Valle R, Putti M, and Siri A S 2008 *Supercond. Sci. Technol.* **21** 095017
[22] Tropeano M, Martinelli A, Palenzona A, Bellingeri E, Galleani dAgliano E, Nguyen T D, Affronte M and Putti M, 2008 *Preprint* 0807.0719 [cond-mat] (in press on Phys. Rev. B)
[23] A. Nucara, A. Perucchi, P. Calvani, T. Aselage and D. Emin, *Phys. Rev. B* 68, 174432 (2003).
[24] Yin Z P, Leb`egue S, Han M J, Neal B, Savrasov S Y and Pickett W E 2008 *Phys. Rev. Lett.* **101** 047001
[25] Trego A L and Machintosh A R 1968 *Physical Review* **166** 495
[26] Elliot R J 1972 *Magnetic properties of rare earth metals* (New York:Plenum Press)
[27] Rizzuto C 1974 *Rep. Prog. Phys.* **37** 147
[28] Sefat A S, McGuire M A, Sales B C, Jin R, Howe J Y and Mandrus D 2008 *Phys. Rev. B* **77** 174503
[29] Dore P, Mirri C, Lupi S, Postorino P, Calvani P, Massidda S, Profeta G to be published
[30] Qazilbash M M, Hamlin J J, Baumbach R E, Maple M B and Basov D N 2008 *Preprint* 0808.3748 [cond-mat]
[31] Li G, Hu W Z, Dong J, Li Z, Zheng P, Chen G F, Luo J L and Wang N L 2008 *Phys. Rev. Lett.* **101** 107004
[32] Dubroka A, Kim K W, M Rössle, Malik V K, Liu R H, Wu G, Chen X H and Bernhard C 2008 *Phys. Rev. Lett.* **101** 097011
[33] Dressel M and Gruner G 2002 *Electrodynamics of Solids* (Cambridge:Cambridge University Press)
[34] Ding H, Richard P, Nakayama K, Sugawara T, Arakane T, Sekiba Y, Takayama A, Souma S, Sato T, Takahashi T, Wang Z, Dai X, Fang Z, Chen G F, Luo J L and Wang N L 2008 *Europhys. Lett.* **83** 47001